\documentclass[abstracton]{scrartcl}
\usepackage{epsfig}
\usepackage{amsbsy,amsmath,amsfonts,amssymb}
\usepackage{graphicx}
\usepackage{float}
\usepackage{psfrag}
\usepackage[latin1]{inputenc}
\usepackage{color}
\usepackage{lineno}
\usepackage{wasysym}
\usepackage{colordvi}
\usepackage[T1]{fontenc}

\def\Journal#1#2#3#4{{#1}~{\bf #2} (#4), #3}

\def\NIMA{{\em Nucl.~Instrum.~Methods}~A}

\def\NPB{{\em Nucl.~Phys.}~B}
\def\PLB{{\em Phys.~Lett.}~B}
\def\PRL{\em Phys.~Rev.~Lett.}
\def\PRD{{\em Phys.~Rev.}~D}

\def\EPJ{{\em Eur.~Phys.~J.}~C}

\newcommand{\Br}{{\text{Br}}}

\newcommand{\pid}{\pi^0_D}

\newcommand{\kpmpipipi}{K^\pm \to \pi^\pm \pi^+ \pi^-}

\newcommand{\kpmpipizpid}{K^\pm \to \pi^\pm \pi^0 \pid}

\newcommand{\kpmpipid}{K^\pm \to \pi^\pm \pid}

\newcommand{\kpmpide}{K^\pm \to \pid e^\pm \nu}

\newcommand{\kpmpidmu}{K^\pm \to \pid \mu^\pm \nu}

\newcommand{\kpmpigg}{K^\pm \to \pi^\pm \gamma \gamma}
\newcommand{\kppigg}{K^+ \to \pi^+ \gamma \gamma}

\newcommand{\kpmpipig}{K^\pm \to \pi^\pm \pi^0 \gamma}

\newcommand{\kpmpipidg}{K^\pm \to \pi^\pm \pid \gamma}

\newcommand{\kpmpipiee}{K^\pm \to \pi^\pm \pi^0 e^+ e^-}

\newcommand{\kpmpieeg}{K^\pm \to \pi^\pm e^+ e^- \gamma}
\newcommand{\kppieeg}{K^+ \to \pi^+ e^+ e^- \gamma}

\newcommand{\kpmpiee}{K^\pm \to \pi^\pm e^+ e^-}

\newcommand{\kpmpipie}{K^\pm \to \pi^+ \pi^- e^\pm \nu (\bar{\nu})}

\newcommand{\kpmpizpide}{K^\pm \to \pi^0 \pid e^\pm \nu (\bar{\nu})}

\newcommand{\meeg}{m_{e^+ e^- \gamma}}

\newcommand{\bdm}{\begin{displaymath}}
\newcommand{\edm}{\end{displaymath}}
\newcommand{\be}{\begin{equation}}
\newcommand{\ee}{\end{equation}}

\begin{document}

%
%
\centerline{EUROPEAN ORGANIZATION FOR NUCLEAR RESEARCH}
\vspace*{1mm}
{\flushright{CERN-PH-EP/2007-033\\ October~4, 2007\\}}
%
\begin{center}
{\bf \huge First Observation and Measurement \\*[4mm] of the Decay \boldmath{${\kpmpieeg}$}}
\end{center}
\begin{center}
\vspace{3mm}
{\Large The NA48/2 Collaboration}\\*[5mm]
%
 J.R.~Batley,
 A.J.~Culling,
 G.~Kalmus,
 C.~Lazzeroni,
 D.J.~Munday,
 M.W.~Slater,
 S.A.~Wotton \\
{\em \small Cavendish Laboratory, University of Cambridge,
Cambridge, CB3 0HE,
U.K.$\,$\footnotemark[1]} \\[0.2cm]
 R.~Arcidiacono,
 G.~Bocquet,
 N.~Cabibbo,
 A.~Ceccucci,
 D.~Cundy$\,$\footnotemark[2],
 V.~Falaleev,
 M.~Fidecaro,
 L.~Gatignon,
 A.~Gonidec,
 W.~Kubischta,
 A.~Norton,
 A.~Maier,
 M.~Patel,
 A.~Peters\\
{\em \small CERN, CH-1211 Gen\`eve 23, Switzerland}
% ...
\\[0.5cm]
%
\end{center}

\setcounter{footnote}{0}
\footnotetext[1]{Funded by the U.K.
Particle Physics and Astronomy Research Council}
\footnotetext[2]{Present address: Istituto di Cosmogeofisica del CNR
di Torino, I-10133 Torino, Italy}
\footnotetext[3]{Present address: Scuola Normale Superiore and INFN,
I-56100 Pisa, Italy}
\footnotetext[4]{Present address: CERN, CH-1211 Gen\`eve 23, Switzerland}
\footnotetext[5]{Also at Centre de Physique des Particules de Marseille,
IN2P3-CNRS, Universit\'e de la M\'editerran\'ee, Marseille, France}
\footnotetext[6] {Also Istituto di Fisica, Universit\`a di Modena,
I-41100 Modena, Italy}
\footnotetext[7]{Istituto di Fisica,
Universit\`a di Urbino, I-61029  Urbino, Italy}
\footnotetext[8]{Present address: Physikalisches Institut,
Universit\"at Bonn, D-53115 Bonn, Germany}
\footnotetext[9]{Funded by the German Federal Minister for Education
and research under contract 05HK1UM1/1}
\footnotetext[10]{Permanent address: SLAC, Stanford University,
Menlo Park, CA 94025, U.S.A.}
\footnotetext[11]{Present address: UCLA,  Los Angeles, CA 90024,
U.S.A.}%
\footnotetext[12]{Funded by the German Federal Minister for Research
and Technology (BMBF) under contract 056SI74}
\footnotetext[13]{Funded by the Austrian Ministry for Traffic and
Research under the contract GZ 616.360/2-IV GZ 616.363/2-VIII, and
by the Fonds f\"ur Wissenschaft und Forschung FWF Nr.~P08929-PHY}

%
%

\begin{abstract}
Using the full data set of the NA48/2 experiment, the decay $\kpmpieeg$ is observed 
for the first time, selecting $120$ candidates with $7.3 \pm 1.7$ estimated background events.
With $\kpmpipid$ as normalisation channel, the branching ratio is determined in a model-independent way to be 
$\Br(\kpmpieeg, m_{ee\gamma} > 260 \: \text{MeV}/c^2)=(1.19 \pm 0.12_\text{stat} \pm 0.04_\text{syst}) \times 10^{-8}$.
This measured value and the spectrum of the $e^+ e^- \gamma$ invariant mass allow a comparison
with predictions of Chiral Perturbation Theory.
\end{abstract}

\vspace*{1mm}

\centerline{\it Accepted by Phys.~Lett.~B}
\vspace*{1mm}

\setcounter{footnote}{0}

%
%


\section{Introduction}

The decay $\kpmpieeg$ is similar to the decay $\kpmpigg$, with one of the photons 
internally converting into a pair of electrons. 
Both decays can be described in the framework of Chiral Perturbation Theory (ChPT). 
The lowest order terms are of order $p^4$, 
where predominantly loop diagrams contribute to the amplitude~\cite{bib:eckerpich}.
This leads to a characteristic signature in the $e^+ e^- \gamma$ invariant mass distribution,
which is favored to be above $2 m_{\pi^+}$ and 
exhibits a cusp at the $2  m_{\pi^+}$ threshold.
In ChPT, the loop contribution is fixed up to a free parameter $\hat{c}$,
which is a function of several strong and weak coupling constants 
and expected to be of ${\cal O}(1)$~\cite{bib:dambrosio}.


Higher order ChPT calculations on $\kppigg$ have been performed, but are model-dependent.
Theoretical predictions for $\kppieeg$ exist~\cite{bib:gabbiani},
following the earlier work on $\kppigg$~\cite{bib:dambrosio}.
The predicted branching ratios lie in the range between
$0.9$ and $1.7\times10^{-8}$, for values of $|\hat{c}| \le 2$.
Experimental results are available only for $\kppigg$, 
based on the observation of 31 signal candidates by the E787 experiment~\cite{bib:e787}.

In this letter, we report the first observation of the decay $\kpmpieeg$
and the model-independent measurement of its branching fraction, using $\kpmpipid$ with $\pid \to e^+ e^- \gamma$
as the normalisation channel.
These results have been derived from the
full data set of the NA48/2 experiment.

\section{Experimental Set-up}
\label{sec:experiment}

The NA48/2 experiment took data in 2003 and 2004 at the CERN SPS.
Two beams of charged particles were produced by a 400~GeV/$c$ proton beam
impinging on a Be target in a 4.8~s long pulse repeated every 16.8~s. 
Positive and negative particles with momenta of $(60\pm 3)$~GeV/$c$ were simultaneously selected by an achromatic
system, which split the two beams in the vertical plane and then recombined them on a common axis.
After passing through a collimator, the beams were split and recombined again 
in a second achromat.
Finally, the two beams passed a cleaning and a defining collimator before entering the
decay volume housed in a 114~m long evacuated tank with a diameter between 1.92 and 2.4~m and 
terminated by a $0.3\%$~radiation lengths thick Kevlar window.
The axes of both beams coincided within 1~mm inside the decay volume. 
The beams were primarily composed of charged pions, with a fraction of $5-6\%$ of $K^\pm$.
On average, about $4.8 \times 10^5$~$K^+$ and $2.7 \times 10^5$~$K^-$ 
per pulse decayed in the fiducial decay volume.
A more detailed description of the beamline can be found in~\cite{bib:asymmetry}.

The decay region was followed by the NA48 detector~\cite{bib:detector}.
The momenta and positions of charged particles were measured in a magnetic spectrometer.
The spectrometer was housed in a helium gas volume and consisted 
of two pairs of drift chambers before and after a dipole magnet with
vertical magnetic field direction, giving a horizontal transverse momentum kick of 120~MeV/$c$.
Each chamber had four views ($x$, $y$, $u$, $v$) with two sense wire planes in each view.
The $u$ and $v$ views were inclined by $\pm 45^\circ$ with respect to the $x$-$y$ plane.
The space points, reconstructed by each chamber, had a resolution of 150~$\mu$m in each projection.
The momentum resolution of the spectrometer in 2003/2004 was measured to be
$\sigma_p/p = 1.02\% \oplus 0.044\% \times p$, with $p$ in GeV/$c$.
The magnetic spectrometer was followed by a segmented plastic scintillator hodoscope 
with one plane of vertical and one plane of horizontal strips, respectively.
It was used to produce fast trigger signals and to provide precise time measurements of charged particles.
The time resolution of the hodoscope was better than 200~ps. 

Photon and electron energies were measured with a 27 radiation length thick liquid-krypton
electromagnetic calorimeter (LKr). 
It was read out longitudinally in 13248 cells of $2\times2$~cm$^2$ cross-section.
The energy resolution was determined to be
$\sigma_E/E = 3.2\%/\sqrt{E} \oplus 9\%/E \oplus 0.42\%$, with $E$ in GeV. 
The spatial and time resolutions were better than 1.3~mm and 300~ps, respectively, 
for photon and electron clusters above 20~GeV.

Additional detector elements, such as the hadron calorimeter 
and the muon and photon veto counters, were not used in the present analysis.

The events were selected by a two-level trigger
which was optimised for events with three charged tracks.
At the first level, three-track events were triggered by requiring coincidences
of hits in the two hodoscope planes. 
The second level trigger was based on the hit coordinates in the drift chambers. It required
at least two tracks to originate from the decay volume with a reconstructed distance of closest approach
of less than 5~cm.
The trigger efficiency for the selection of the normalisation channel $\kpmpipid$
was $(96.48 \pm 0.05)\%$, determined from data events taken with a down-scaled 
complementary trigger.

In total, NA48/2 collected about $18 \times 10^9$ triggers. 
In the course of data-taking, the magnet polarities of both the beamline and the 
spectrometer were regularly reversed to have similar conditions for decays of positive and negative kaons.

\section{Monte Carlo Simulation}

In order to compute the acceptance of signal, normalisation, and background channels, a detailed
GEANT-based~\cite{bib:geant} Monte Carlo simulation was employed, which included
the full detector and material description, stray magnetic fields, drift chamber
inefficiencies and misalignment, and beamline simulation.

For the signal channel, the full matrix element was used~\cite{bib:gabbiani},
with a value of $\hat{c}=1.8$, in agreement with the measurement of $\kppigg$~\cite{bib:e787}.

For all other channels, if not otherwise mentioned below, the known theoretical
matrix elements were used. Radiative corrections were
applied to the simulation of signal, normalisation, and $\kpmpiee$ by using the PHOTOS package~\cite{bib:photos}.

\section{Data Selection}

The analysis described here is based on the full data set of the NA48/2 experiment,
recorded in 2003 and 2004. 
The selection of the signal events was performed in two steps.
At first, described in the next section, a set of basic selection criteria was applied
to define the signal region and to assure the quality of the selected candidate events.
In a second step, described in the following section, further 
selection criteria were applied to suppress contributions of the various background 
sources.

\subsection{Event Selection}

Each selected event had to have at least one combination of
three tracks with a total charge of $\pm 1$
and one cluster in the LKr calorimeter not associated with any track.
Each track was required to have a radial distance
from the detector axis of at least 12~cm in the first drift chamber 
and to lie well inside the active LKr calorimeter region, i.e.\ 
well inside its outer edge and more than 15~cm from the detector axis.
The distance between any two tracks in the first drift chamber had to exceed 2~cm.
This latter requirement rejects all events with external photon conversions in the
detector material before the spectrometer.

The three tracks had to be compatible with originating from the same decay vertex and to have
a distance of closest approach of less than 4~cm for each of the three pairs
of two tracks. 
The longitudinal position of the reconstructed decay vertex had to be more than 2~m 
and less than 98~m down-stream of the final collimator and within a radius of 3~cm around the beam axis.
The track times, measured in the scintillator hodoscope, had to be at most $\pm 3$~ns from the mean of the track times.
In less than $1\%$ of the events, no hodoscope information was available for at least one track. 
For those events, the track times measured in the drift chambers were taken and required to be at most $\pm 6$~ns 
from the mean track time.

Pions and electrons were identified by the ratio $E/p$ of energy deposited in the LKr calorimeter
and momentum measured in the spectrometer. Electrons and pions were required to have
$E/p > 0.94$ and $E/p<0.8$, respectively. With these requirements the probability
for mis-identification of electrons or pions is of the order of a few per mille.
Two tracks of opposite charge had to be identified as electrons with each having a momentum
greater than 3~GeV/$c$. The third track had to be a pion candidate
and had to have a momentum greater than 4~GeV/$c$.

Photon candidates were defined as calorimeter clusters unassociated to charged tracks 
and required to lie 15~cm from the detector axis and well inside the outer edge of the LKr.
The distance to the projected impact point of any pion candidate had to exceed 25~cm,
the distance to the electron tracks or any other possible cluster had to exceed 10~cm.
The reconstructed cluster energy had to be greater than 3~GeV, and the time difference to the mean
of the cluster time and the track times measured in the drift chambers had to be smaller than 6~ns.

The sum of pion, electron, positron, and photon momenta was required
to be between 54 and 66~GeV/$c$. 
An energy centre-of-gravity
$(x_\text{cog}, y_\text{cog}) = ( \sum_i x_i E_i \, / \, \sum_i E_i, \sum_i y_i E_i \, / \, \sum_i E_i )$
was defined by using the transverse positions $x_i$ and $y_i$ 
and the energies $E_i$ of the projected tracks and the photon cluster 
at the front surface of the LKr calorimeter.
The tracks were projected onto the LKr surface from their positions and directions in the first drift chamber
before the spectrometer magnet.
The radial distance between the energy centre-of-gravity and the beam had to be less than 3~cm.

To suppress background events coming from decays with more than one photon,
we required that no other unassociated cluster was in-time with the event.
Since this would reject also events with photons from bremsstrahlung 
on detector material or with shower fluctuations of pion showers, 
we still allowed additional clusters with $E \, [\text{GeV}] < 7 - 0.14 \, d_e \, [\text{cm}]$ or
$E\, [\text{GeV}] < 15 - 0.25 \, d_\pi\, [\text{cm}]$, where $d_e$ and $d_\pi$ are the distances of the cluster to the
impact point of an electron and the pion track, respectively. This requirement 
against additional unassociated clusters rejected about $0.3\%$ of all events.


With these basic selection criteria, a sample of about 22.8~million events was obtained.
At this level of the selection, the data were dominated by $\kpmpipid$ decays, where the $\pi^0$
underwent a Dalitz decay $\pid \to e^+ e^- \gamma$.


\subsection{Background Suppression}
\label{sec:bkgsuppression}

A number of additional selection criteria had to be applied to effectively suppress
the remaining background.
In the following, we examine the possible sources of background to the signal.

\begin{itemize}

\item{\sfb \boldmath $\kpmpipid$, $\kpmpide$, and $\kpmpidmu$ decays}

The decay $\kpmpipid$ ($K_{2\pi D}$) with $\pid \to e^+ e^- \gamma$
has exactly the same signature as the signal channel.
We therefore rejected events for which 
$120$~MeV$/c^2 < \meeg < 150$~MeV$/c^2$. To evaluate this requirement,
we assigned the electron mass to each track and applied the cut to both opposite-charged track combinations.
This completely rejects also the semileptonic decays $\kpmpide$ and $\kpmpidmu$ 
as well as the small amount of doubly-misidentified $K_{2\pi D}$ events, where both
the pion and an electron are misidentified. 

\item{\sfb \boldmath $\kpmpipidg$ decays}

The decay $\kpmpipidg$ 
consists of two amplitudes: Inner Bremsstrahlung (IB)
and Direct Emission (DE). If the radiative photon is lost, the decay is rejected
by the cut against $\kpmpipid$ decays. However, if the photon of the $\pid$ decay 
is lost, the decay may fake a signal event.
$\kpmpipidg$ events are the major background source and contribute with $3.1\pm0.5$ IB and $0.12\pm0.03$ DE events,
as determined from the simulation, using the measured rates of IB and DE transitions~\cite{bib:pdg}.

\item{\sfb \boldmath $\kpmpipiee$ decays}

The decay $\kpmpipiee$, which comes from $\kpmpipig$ with an internal conversion of the 
additional photon, has not been measured yet. 
By evaluating the internal conversion probability, we estimated
its branching fraction to be half of that of $\kpmpipidg$. Due to the uncertainty of the estimation, we assigned a 
$\pm 50\%$ systematic uncertainty to this value.
$\kpmpipiee$ events were simulated by modifying the $\kpmpipig$ simulation;
the conversion of the photon was added by generating the photon mass with a 
probability density proportional to the inverse square of the photon mass.
The ratio of IB and DE amplitudes was taken from $\kpmpipig$ decays~\cite{bib:pdg}.
From this, we estimated the amount of background from $\kpmpipiee$ to
$1.6\pm0.5\pm0.7$ events, where the second uncertainty comes from the estimation of the branching fraction.
The contribution from DE is practically negligible.


\item{\sfb \boldmath Radiative $\kpmpiee$ decays}

The probability of the rare decay $\kpmpiee$ with an observable $\gamma$ from internal or external brems\-strahlung is
of ${\mathcal O}(10^{-8})$, similar to the signal channel. 
To reject these events, we made use of the different decay kinematics in the 
$e^+ e^- \gamma$ rest frame.
For the signal, the photon repels from the $e^+ e^-$ system, while in case of
$\kpmpiee + \gamma_\text{brems}$ the electrons, in the $e^+ e^- \gamma$ rest frame, fly back-to-back, and the photon is close to one of them. 
We therefore required $\varangle(e^+,e^-) < \min(\varangle(e^\pm,\gamma))$
for the angles between $e^+$ and $e^-$ and between $e^\pm$ and the photon, respectively, in the $e^+ e^- \gamma$ rest frame
(see Figure~\ref{fig:pieecut}).
The remaining background from $\kpmpiee$ decays was determined to be $0.8\pm0.5$ events 
from MC simulation. 
This estimate includes a systematic error of $\pm 50\%$, which reflects a disagreement
between $\kpmpiee$ data and MC in event numbers in the region with
$\varangle(e^+,e^-) > \min(\varangle(e^\pm,\gamma))$.


\begin{figure}[t]
  \begin{center}
    \epsfig{file=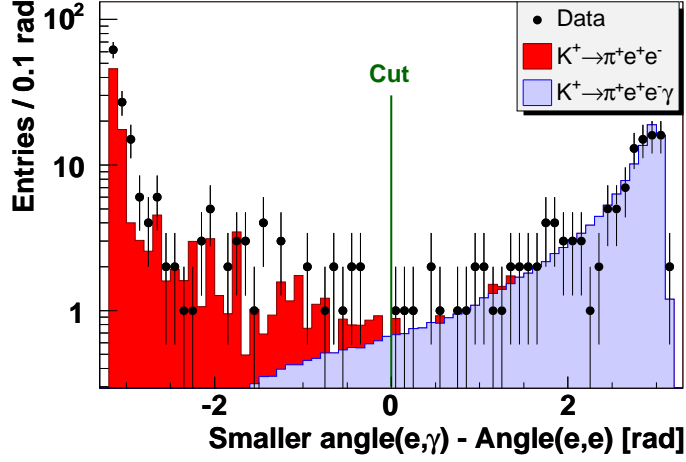,width=0.68\textwidth}
  \end{center}
  \caption{Difference of the smaller of the angles between the photon and $e^\pm$ and
           the angle between $e^+$ and $e^-$ in the $e^+ e^- \gamma$ rest frame for data (circles)
           and signal and radiative $\kpmpiee$ MC simulation.}
  \label{fig:pieecut}
\end{figure}

\item{\sfb \boldmath $\kpmpipizpid$ decays}

The decay $\kpmpipizpid$ was strongly suppressed by the cut on additional clusters and the
rejection of the $\pid$ decays. 
Its contribution to the signal region was estimated by MC simulation to be $0.7\pm0.7$ events.

\item{\sfb Accidental activity}

Accidental overlap of separate events may fake signal events.
To estimate the amount of such events in the signal sample, we studied
the sidebands of the time distributions in the hodoscope and the LKr calorimeter. 
One event was found in the calorimeter time sideband, which corresponds to a background estimation of $1 \pm 1$~events
from accidental activity.

\end{itemize}

Other potential sources of background as e.g.\ $\kpmpipipi (\gamma)$, $\kpmpipie$, or $\kpmpizpide$ 
were found to be irrelevant.

The signal region was defined by requiring $480 < m_{\pi^\pm e^+ e^- \gamma} < 505$~MeV/$c^2$.
Since ChPT predicts only small signal rate 
and the background increases for low values of $\meeg$, we also required $\meeg > 260$~MeV/$c^2$.
120 signal candidates were found, including an estimated 
total background of $7.3 \pm 1.7$ events.
The background channels are listed in Table~\ref{tab:backgrounds} together with their
respective branching fractions and expected contributions to the signal.
All background expectations were obtained by normalising to the total kaon flux.

\begin{figure}[htbp]
  \begin{center}
    \epsfig{file=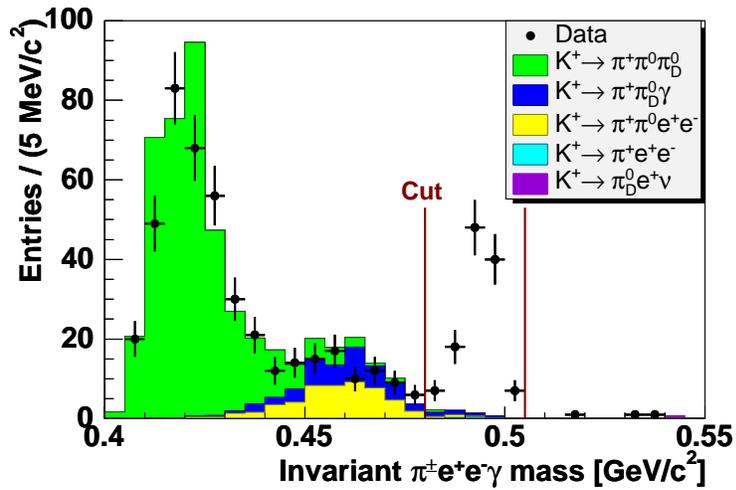,width=0.7\textwidth}
    \epsfig{file=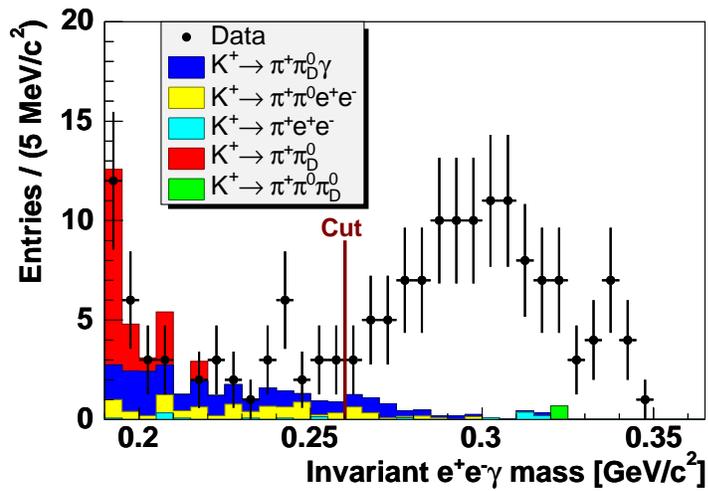,width=0.7\textwidth}
  \end{center}
  \caption{Selected signal candidates and background expectation from MC simulation:
           (top) $\pi^\pm e^+ e^- \gamma$ invariant mass, 
           (bottom) $e^+ e^- \gamma$ invariant mass.
           The vertical lines indicate the accepted region for the branching ratio measurement.}
  \label{fig:signal_1d}
\end{figure}

The signal acceptance, as determined from MC simulation, depends on $\meeg$. It was between 6\% and 7\%
for $260 < \meeg < 330$~MeV/$c^2$, and decreased to $2.5\%$ for events near the kinematical edge.
The projections of the signal candidates on $m_{\pi^\pm e^+ e^- \gamma}$ and $\meeg$, together with
the background contributions, are shown in Figure~\ref{fig:signal_1d}.

\begin{table}[htbp]
  \begin{center}
    \begin{tabular}{lcc}
      \hline \hline
      Background source & Branching ratio           & Expected events   \\
      \hline
      $\kpmpipidg$ (IB) & $3.3 \times 10^{-6}$      &  $3.1 \pm 0.5$    \\
      $\kpmpipidg$ (DE) & $5.3 \times 10^{-8}$      &  $0.12 \pm 0.03$  \\
      $\kpmpipiee$ (IB) & $\sim 1.7 \times 10^{-6}$ &  $1.6 \pm 0.9$    \\
      $\kpmpipiee$ (DE) & $\sim 2.6 \times 10^{-8}$ &  $0.02 \pm 0.01$  \\
      $\kpmpiee$        & $2.9 \times 10^{-7}$      &  $0.8 \pm 0.5$    \\
      $\kpmpipizpid$    & $2.1 \times 10^{-4}$      &  $0.7 \pm 0.7$    \\
      Accidentals       & ---                       &  $1.0 \pm 1.0$    \\
      \hline
      Sum               &                           &  $7.3 \pm 1.7$   \\
      \hline \hline
    \end{tabular}
    \caption{Relevant background sources and number of expected events
             in the signal region with all selection criteria applied.
             Except for $\kpmpipiee$, the branching fractions are taken from~\cite{bib:pdg};
             for $\kpmpipidg$ and $\kpmpipiee$ they are defined for $T_\pi = 55 - 90$~MeV.}
    \label{tab:backgrounds}
  \end{center}
\vspace*{-5mm}
\end{table}


\subsection{Normalisation Channel}
\label{sec:normalisation}

For the normalisation channel $\kpmpipid$ exactly the same selection as for the signal was applied,
but without the criteria on the $e^+ e^- \gamma$ invariant mass and the $e^+ e^- \gamma$ decay angles.
Instead, the $e^+ e^- \gamma$ invariant mass was required to be within
$m_{\pi^0} - 35$~MeV/$c^2$ and $m_{\pi^0} + 30$~MeV/$c^2$. The asymmetry of this cut takes into account
the radiative tail in the $m_{e^+ e^- \gamma}$ distribution.
We found about 18.7~million $K_{2\pi D}$ candidates including an estimated background of about $0.5\%$.
The acceptance of the selection was determined to be $5.0\%$ from MC simulation.
The branching fraction of the normalisation is
$\Br(\kpmpipid, \pid \to e^+ e^- \gamma) = (2.51 \pm 0.07) \times 10^{-3}$~\cite{bib:pdg}.
From this, we determined the total flux of kaon decays in the fiducial volume 
to be $\Phi_K = (1.48 \pm 0.04) \times 10^{11}$.

\section{Results}


To determine the branching fraction in a model-independent way, we computed
a partial branching fraction for each 5~MeV/$c^2$ wide $m_{e^+e^-\gamma}$ interval $i$ from
\[
\Br_{\;\!i\;\!}(\kpmpieeg) \; = \; \frac{N_i^{\pi e e \gamma} - N_i^\text{bkg}}{A_i^{\pi e e \gamma} \cdot \epsilon} \times \frac{1}{\Phi_K},
\]
with the numbers $N_i^{\pi e e \gamma}$ and $N_i^\text{bkg}$ of observed signal and estimated backgrounds events,
and the signal acceptance $A_i^{\pi e e \gamma}$ in bin $i$. The overall trigger efficiency is $\epsilon$ and $\Phi_K$ the total kaon flux.
By summing over the bins above $m_{e^+e^-\gamma} = 260$~MeV/$c^2$, we obtained
$\Br(\kpmpieeg, m_{e^+e^-\gamma} > 260 \: \text{MeV}/c^2) = (1.19 \pm 0.12_\text{stat}) \times 10^{-8}$,
where the error is from data statistics only.
This result is independent of the value of $\hat{c}$ 
and any theoretical assumption of the $\meeg$ distribution.


Several potential sources of systematic errors 
can affect the result and have been studied.

The background estimation has a total uncertainty of {$\pm 1.7$}~events, as explained before,
which results in an uncertainty of $\pm 1.5\%$ on the result.


Possible imperfections of the description of the detector acceptance in the Monte Carlo simulation
might also cause systematic effects on the branching fraction measurement.
For an estimation of such effects we have varied the main selection cuts.
To not fall victim of statistical fluctuations in the signal channel, 
the variations have only been performed in the normalization channel.
This leads to a conservative estimate, since detector systematics are 
expected to cancel between signal
and normalization. We found maximum changes of the result of the order of $\pm 0.4\%$,
which we assigned as the systematic uncertainty due to the detector acceptance.

The particle identification via the $E/p$ ratio could not be perfectly
modelled in the simulation. However, the inefficiencies are expected 
to almost completely cancel between signal and normalization mode.
The residual uncertainty on the electron identification was measured
from $K^0_{e3}$ decays to better than $0.1\%$. The uncertainty of the
pion identification efficiency was determined from variations of the $E/p$ criterion
in the normalization channel to be at most $\pm 0.3\%$.
Combining both, we assigned a total uncertainty of $\pm 0.4\%$ due to particle identification.

The overall trigger efficiency should be the same for signal and normalisation to a great extent.
A difference could only arise from the slightly different event topologies.
Due to the lack of statistics, the trigger efficiency could not be measured for signal events. 
We therefore studied the dependency of the trigger efficiency of $\kpmpipid$ events as a function of the event topology.
From this, we obtained a systematic uncertainty of $\pm 0.6\%$.

The statistical error of the signal and normalisation MC samples contributes to $\pm 0.9\%$.

Finally, the external inputs of $\Br(\kpmpipid)$ and $\Br(\pid \to e^+ e^- \gamma)$
add an uncertainty of $\pm 2.7\%$~\cite{bib:pdg}.
This is identical with the error quoted on the kaon flux in Section~\ref{sec:normalisation}.

All uncertainties of the measurement are listed in Table~\ref{tab:systsum}.
The final result on the branching ratio is
\[
\Br(\kpmpieeg, m_{e^+e^-\gamma} > 260 \: \text{MeV}/c^2) \; = \; (1.19 \pm 0.12_\text{stat} \pm 0.04_\text{syst}) \times 10^{-8}.
\]

\begin{table}[tbp]
  \begin{center}
    \begin{tabular}{lcc}
      \hline \hline
      Source                       & $\Delta \Br/\Br$ & $\Delta \Br$ $[10^{-8}]$  \\*[1mm] \hline
      Background subtraction       & $\pm \: 1.5\%$          & $\pm \: 0.017$ \\
      Electron/pion identification & $\pm \: 0.4\%$          & $\pm \: 0.005$ \\
      Detector acceptance          & $\pm \: 0.4\%$          & $\pm \: 0.005$ \\
      Trigger efficiency           & $\pm \: 0.6\%$          & $\pm \: 0.007$ \\
      MC statistics                & $\pm \: 0.9\%$          & $\pm \: 0.011$ \\
      Normalisation                & $\pm \: 2.7\%$          & $\pm \: 0.032$ \\ \hline
      Total systematic uncertainty & $\pm \: 3.3\%$          & $\pm \: 0.04$ \\ \hline 
      Statistical uncertainty      & $\pm \: 9.7\%$          & $\pm \: 0.12$ \\ \hline \hline
    \end{tabular}
  \end{center}
  \caption{Summary of uncertainties of the branching ratio measurement.}
  \label{tab:systsum}
  \vspace*{3mm}
\end{table}

\begin{figure}[htbp]
  \begin{center}
    \epsfig{file=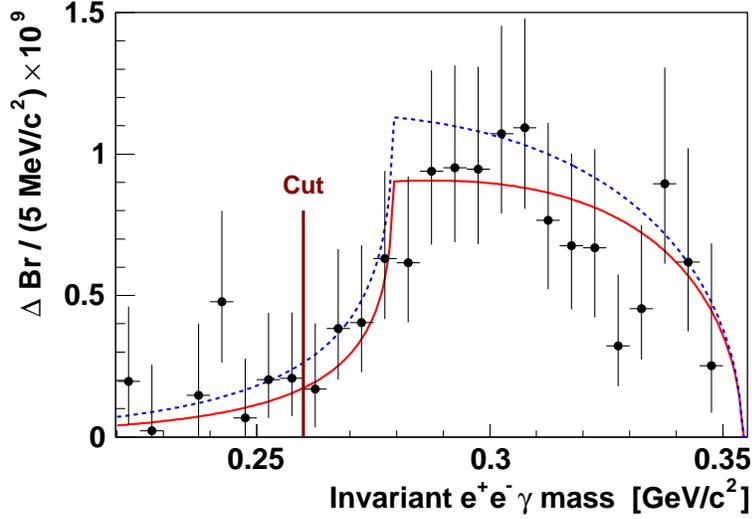,width=0.7\textwidth}
  \end{center}
  \caption{Partial $\kpmpieeg$ branching fractions as function of the $e^+ e^- \gamma$ invariant mass. 
           The signal region is defined for $\meeg > 260$~MeV/$c^2$.
           The lines are the expectations from ref.~\cite{bib:gabbiani}
           for $\hat{c} = 0.90$ (solid, best fit)
           and $\hat{c} = 1.8$ (dashed, estimate from $\kppigg$~\cite{bib:e787}).}
  \label{fig:eegspectrum}
\end{figure}

The distribution of the partial branching fractions is shown in Figure~\ref{fig:eegspectrum} and tabulated
in Table~\ref{tab:eegspectrum}. 
The quoted errors are confidence intervals for the unknown true value.
We chose Pearson's $\chi^2$ intervals for Poisson statistics~\cite{bib:cowan} for them,
defined as $\sigma_{\pm} = \sqrt{n_i + 1/4} \pm 1/2$ for each data bin with entry $n_i$,
before background subtraction, acceptance correction, and normalisation.
The uncertainties on the background and acceptance estimates were added in quadrature in each bin, while the
global systematic uncertainties --- dominated by the normalisation --- 
are not quoted in Figure~\ref{fig:eegspectrum} and Table~\ref{tab:eegspectrum}.

\begin{table}[htbp]
  \begin{center}
  \renewcommand{\arraystretch}{1.4}
  \begin{tabular}{cc}
   \hline \hline
   $m_{e^+e^-\gamma}$ interval & $\Delta \, \Br$ $[10^{-9}]$ \\ \hline
    260 -- 265 MeV/$c^2$ & $0.17 \, {+0.23 \atop -0.13}$   \\
    265 -- 270 MeV/$c^2$ & $0.38 \, {+0.28 \atop -0.18}$   \\
    270 -- 275 MeV/$c^2$ & $0.40 \, {+0.27 \atop -0.18}$   \\
    275 -- 285 MeV/$c^2$ & $0.63 \, {+0.31 \atop -0.21}$   \\
    280 -- 280 MeV/$c^2$ & $0.62 \, {+0.30 \atop -0.21}$   \\
    285 -- 295 MeV/$c^2$ & $0.94 \, {+0.36 \atop -0.26}$   \\
    290 -- 290 MeV/$c^2$ & $0.95 \, {+0.36 \atop -0.26}$   \\
    295 -- 300 MeV/$c^2$ & $0.95 \, {+0.36 \atop -0.26}$   \\
    300 -- 305 MeV/$c^2$ & $1.07 \, {+0.38 \atop -0.28}$   \\
   \hline \hline
  \end{tabular}
  \hspace*{9mm}
  \begin{tabular}{cc}
   \hline \hline
   $m_{e^+e^-\gamma}$ interval & $\Delta \, \Br$ $[10^{-9}]$ \\ \hline
    305 -- 310 MeV/$c^2$ & $1.09 \, {+0.39 \atop -0.29}$    \\
    310 -- 315 MeV/$c^2$ & $0.77 \, {+0.34 \atop -0.24}$    \\
    315 -- 320 MeV/$c^2$ & $0.68 \, {+0.33 \atop -0.23}$    \\
    320 -- 325 MeV/$c^2$ & $0.67 \, {+0.35 \atop -0.25}$    \\
    325 -- 330 MeV/$c^2$ & $0.32 \, {+0.25 \atop -0.14}$    \\
    330 -- 335 MeV/$c^2$ & $0.45 \, {+0.29 \atop -0.18}$    \\
    335 -- 340 MeV/$c^2$ & $0.89 \, {+0.41 \atop -0.28}$    \\
    340 -- 345 MeV/$c^2$ & $0.62 \, {+0.40 \atop -0.25}$    \\
    345 -- 350 MeV/$c^2$ & $0.25 \, {+0.43 \atop -0.17}$    \\
   \hline \hline
  \end{tabular}
  \renewcommand{\arraystretch}{1.0}
  \end{center}
  \caption{Partial $\kpmpieeg$ branching fractions dependent on the $e^+e^-\gamma$ invariant mass.
           Quoted are uncertainties from data and MC statistics and background estimation.
           All other uncertainties are completely correlated and 
           amount to $\pm 3.2\%$, dominated by the normalisation.}
  \label{tab:eegspectrum}
\end{table}

We used the measured branching fraction and the shape of the 
$e^+ e^- \gamma$ spectrum to extract a value for the parameter $\hat{c}$.
Performing a least squares fit of the absolute prediction given in ref.~\cite{bib:gabbiani} 
to the data with $\meeg > 260$~MeV/$c^2$, 
we obtained $\hat{c} = 0.90 \pm 0.45$, where the error is dominated by the data statistics. 
The quality of the fit was $\chi^2/n_\text{dof} = 8.1/17$.
This result is in agreement within about 1.2 standard deviations with the value of $1.8 \pm 0.6$,
previously measured in $\kppigg$~\cite{bib:e787}, and has a somewhat smaller error.
Figure~\ref{fig:eegspectrum} shows the predicted spectrum 
for our best fit value of $\hat{c}$ and the previously found value,
together with the background and acceptance corrected data.

Using our measured value of $\hat{c}$ and ref.~\cite{bib:gabbiani}, we computed 
the differential branching fraction for $\meeg < 260$~MeV/$c^2$ and added it to our measured result.
We then obtained for the total branching fraction
$\Br(\kpmpieeg) \; = \; (1.29 \pm 0.13_\text{exp} \pm 0.03_{\hat{c}}) \times 10^{-8}$,
where the first uncertainty is the combined statistical and systematic error, 
and the second reflects the uncertainty in $\hat{c}$.

\section{Acknowledgements}

It is a pleasure to thank the technical staff of the participating
laboratories, universities, and affiliated computing centres for their
efforts in the construction of the NA48 apparatus, in the
operation of the experiment, and in the processing of the data.
We are grateful to F.~Gabbiani for providing us the code 
to compute the $\kpmpieeg$ matrix element.

%
%


\begin{thebibliography}{00}

\bibitem{bib:eckerpich}     G.~Ecker, A.~Pich, and E.~de~Rafael,
                            \Journal{\NPB}{303}{665}{1988}.

\bibitem{bib:dambrosio}     G.~D'Ambrosio and J.~Portol\'es,
                            \Journal{\PLB}{386}{403}{1996}.

\bibitem{bib:gabbiani}      F.~Gabbiani, 
                            \Journal{\PRD}{59}{094022}{1999}.

\bibitem{bib:e787}          P.~Kitching {\em et al.} (E787 Collaboration), 
                            \Journal{\PRL}{79}{4079}{1997}.

\bibitem{bib:asymmetry}     J.R.~Batley {\it et al.} (NA48/2 Collaboration),
                            CERN-PH-EP-2007-021 \\ (arXiv:0707.0697), submitted to \EPJ.

\bibitem{bib:detector}      V.~Fanti {\it et al.} (NA48 Collaboration),
                            \Journal{\NIMA}{574}{433}{2007}.

\bibitem{bib:geant}         GEANT Detector Description and Simulation Tool,
                            CERN Program Library Long Write-up W5013 (1994).

\bibitem{bib:photos}        E.~Barberio and Z.~Was,
                            \Journal{Comp.~Phys.~Comm.}{79}{291}{1994}.

\bibitem{bib:pdg}           W.M.~Yao {\it et al.} (Particle Data Group),
                            \Journal{J.~Phys.}{G 33}{1}{2006}.

\bibitem{bib:cowan}         G.~Cowan, ``Statistical Data Analysis'',
                            Oxford University Press, Oxford, 1998.

\end{thebibliography}
\end{document}